# Low-Loss Optical Nanofibers with Submicron Waist Diameters and Millimeter-Scale Waist Lengths

GUANGHUI SU, TIMOTHY H. NGUYEN, BALTHAZAR LOGLIA, AARON WEINSTEIN[1], HANBO YANG[2], NAMI UCHIDA[3], MARIAM MCHEDLIDZE, XUEJIAN WU*

*Department of Physics, Rutgers University-Newark, Newark, NJ 07102, USA*
[1]*Present address: Union County Magnet High School, Scotch Plains, NJ 07076, USA*
[2]*Present address: Homer L. Dodge Department of Physics and Astronomy, The University of Oklahoma, Norman, OK 73019, USA*
[3]*Present address: Wyant College of Optical Sciences, The University of Arizona, Tucson, AZ 85721, USA*
*\*xuejian.wu@rutgers.edu*

**Abstract:** Optical nanofibers with subwavelength diameters generate strong evanescent fields, enabling efficient light–matter interactions for optical sensing, spectroscopy, and cold-atom experiments. We report a heat-and-pull system for fabricating low-loss optical nanofibers with controllable waist dimensions and investigate the fabrication limits for achieving small waist diameters and long waist lengths. We study factors that influence fabrication performance, including flame geometry, nanofiber dimensions, and surface contamination. Using a multi-hole torch tip that provides a relatively large and uniform heating region, we achieve reproducible fabrication with optical transmission above 99.9% for waist diameters as small as 200 nm for a 1-mm waist length and 250 nm for a 50-mm waist length. We also develop a preparation procedure for fiber splicing and cleaning to minimize transmission loss caused by surface contamination. In addition, we measure long-term transmission degradation due to dust accumulation in a typical laboratory environment and find that nanofibers fabricated in an enclosed setup maintain transmission above 85% for more than 1 hour for nanofibers with a 300-nm waist diameter and waist lengths ranging from 1 to 30 mm. Our work provides practical guidelines for constructing nanofiber fabrication platforms and producing low-loss nanofibers for optics and atomic physics applications.

## 1. Introduction

Optical nanofibers, fabricated by tapering single-mode optical fibers to subwavelength diameters, have emerged as important tools in optical sensing [1], spectroscopy [2], and the manipulation of microscopic particles and cold atoms [3]. When the nanofiber waist becomes thinner than the optical wavelength, the fiber not only guides light but also generates a tightly confined evanescent field around the waist. By modeling an optical nanofiber as a cylindrical clad-air waveguide with a step-index profile, the guided and evanescent-field modes can be numerically evaluated [4]. Although high-order modes appear as the waist diameter decreases, only the fundamental mode remains when the waist radius satisfies the single-mode condition $a \leq 2.405\lambda/(2\pi(n_1^2-n_2^2)^{1/2})$, where a is the radius of the waist, $\lambda$ is the optical wavelength, $n_1$ is the refractive index of the cladding, and $n_2$ is that of air [4]. For an optical wavelength of 780 nm, the single-mode cutoff diameter for a silica nanofiber is approximately 570 nm. At this diameter, about 15% of the optical power resides in the evanescent field, and the optical mode extends beyond the fiber boundary, corresponding to an effective mode diameter on the order of 600-700 nm [4].

Optical transmission loss in optical nanofibers primarily arises from scattering and mode-coupling in both the taper regions and the waist [5-10]. State-of-the-art low-loss optical nanofibers are typically fabricated using heat-and-pull processing of single-mode optical fibers, enabling high waist uniformity and minimal surface roughness [10, 11]. To ensure sufficient viscosity without thermal damage to silica, the heating temperature is typically set to around 1500 °C. Various heating sources have been employed, including flame burners [12-14], electric heaters [15–17], intense laser beams [18, 19], and plasmonic heating elements [20]. Among these approaches, oxyhydrogen flames, which provide a clean, localized, and uniform heated region, remain one of the most widely used and reliable methods for fabricating low-loss optical nanofibers [21-31]. As the flame is scanned along the optical fiber during pulling, the heat-and-pull fabrication process is also known as the flame-brushing technique.

To date, optical nanofibers and their fabrication systems are not commercially available. Research groups construct their own heat-and-pull platforms to meet specific experimental requirements. This need is further reinforced by the fragility of optical nanofibers, whose transmission performance deteriorates over time due to environmental contamination or fails entirely due to breakage during handling [32-34]. Moreover, many experiments demand optical nanofibers with precisely tailored specifications, such as custom-designed waist and taper dimensions, which can only be achieved through in-house fabrication [35-40]. Therefore, establishing a robust and reproducible fabrication system is essential for conducting experiments with optical nanofibers.

Here, we present a comprehensive procedure for constructing an optical nanofiber fabrication system, including both hardware and software components. Furthermore, we perform extensive experimental characterizations, including the influence of flame geometry, the measurements of optical transmission as a function of nanofiber waist diameter and length, and the effect of dust contamination before and after fabrication. Our results provide practical guidelines for building optical nanofiber fabrication systems and reproducibly fabricating low-loss optical nanofibers with waist diameters as small as 200 nm and waist lengths ranging from 1 to 50 mm.

## 2. Optical Nanofiber Fabrication System

### 2.1 Principle of Heat-and-Pull

The heat-and-pull method fabricates optical nanofibers by softening a single-mode optical fiber and stretching it from both ends. As illustrated in Fig. 1 (a), a small flame creates a localized hot zone around the fiber, and the two motorized stages pull the fiber symmetrically at a controlled relative velocity. Within the hot zone, silica becomes sufficiently viscous that the applied tensile force reduces the fiber diameter while conserving material volume.

For a stationary flame with an effective size of $L_0$, the optical fiber diameter $D(z)$ at the taper length $z$ follows an exponential tapering law governed by volume conservation $D(z)=D_0\exp(-(v_f/2)t/L_0)$, where $D_0$ is the initial optical fiber diameter, $t$ is the flame heating duration, and $v_f$ is the effective pulling velocity [12, 21, 22, 25]. Here, $v_f$ is defined as the relative velocity between two symmetrically moving translation stages, each pulling the optical fiber at a constant velocity of $v_f/2$. The taper length $z$ along one axial direction is equal to the axial elongation $x/2$ imposed by a single translational stage, where $z=x/2=(v_f/2)t$.

To control the taper profile, the flame is scanned along the fiber axis during pulling. In this configuration, the optical fiber is sequentially heated over a spatial range determined by the flame-scanning amplitude $L(x)$. As a result, the taper fabricated along one axial direction is given by $z=x/2+\Delta L(x)/2$, where x denotes the fiber elongation imposed by the translation stages and $\Delta L(x)$ represents the additional extension arising from the variation of flame-scanning amplitude. Based on volume conservation, the resulting taper geometry $D(z)$ and the corresponding axial length $z$ can be calculated as a function of $x$, where the elongation $x$ is determined by the fabrication time $t$ [12, 21, 22, 25].

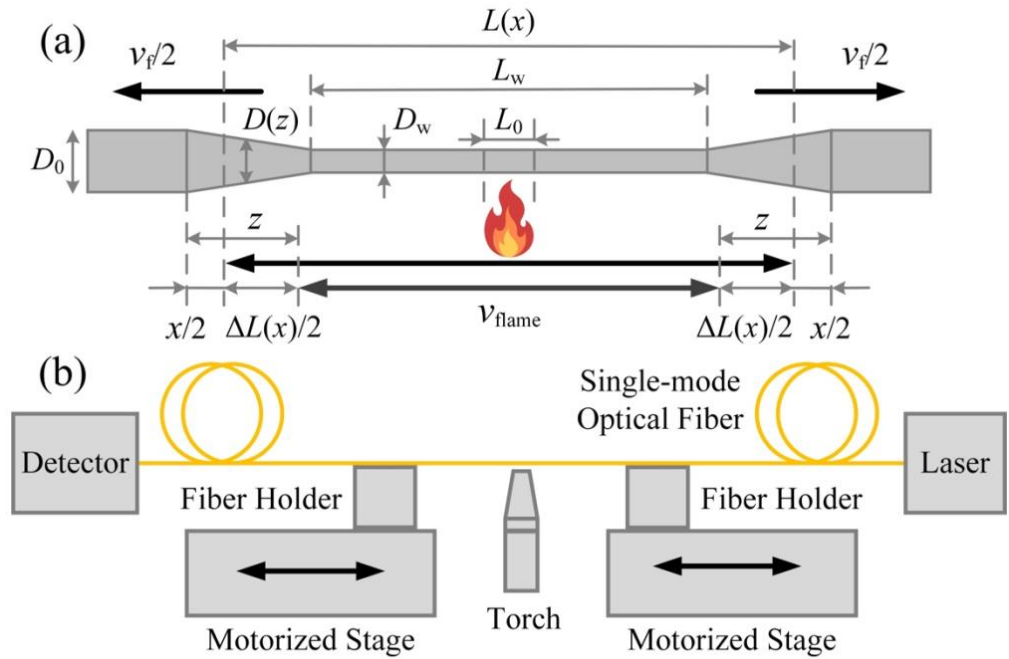

Fig. 1. Principle of the heat-and-pull fabrication process. (a) Schematic of flame scanning during fiber elongation. (b) Schematic of the experimental setup.

To avoid mechanically moving the flame during fabrication, the flame-scanning function is commonly implemented using the translation stages that pull the optical fiber. As shown in Fig. 1 (b), a single-mode optical fiber is mounted on two fiber holders, each attached to a linear translation stage, so that the relative motion between the fiber and the stationary flame defines the flame-scanning amplitude. Additionally, a transmitted laser beam is typically used to monitor the cutoffs of higher-order optical modes during fabrication, enabling measurement of optical transmission loss and providing an indication of the nanofiber waist diameter.

## 2.2 Apparatus

Figure 2 shows the heat-and-pull apparatus assembled on an aluminum optical breadboard (PBG12110, Thorlabs). Two pairs of optical fiber holders (HFF003, Thorlabs) are mounted on ultra-precision motorized linear stages (XMS100-S, Newport) to secure a single-mode optical fiber (SM800-5.6-125, Thorlabs) on both ends. To align the V-grooves of the optical fiber holders, one pair is mounted on a manual 3-axis stage (MBT616D, Thorlabs), while the other pair is fixed on a stationary stage (AMA034, Thorlabs).

Two digital microscopes are employed to monitor optical fiber alignment during mounting and fabrication. Each digital microscope consists of a long-working-distance objective (10X Plan Apo, Mitutoyo), a tube lens (MT-4, Edmund Optics), a 152.5-mm extension tube (56-992, Edmund Optics), and a camera (CS165MU1, Thorlabs). One microscope is mounted on a motorized 2-axis stage (PT1-Z9, Thorlabs) and provides a front-view of the optical fiber, enabling measurement of the vertical separation between the optical fiber and the torch tip. A second digital microscope is used to observe the optical fiber from the top view, facilitating alignment of the optical fiber with respect to the flame. A free-running 852-nm distributed Bragg reflector laser diode (852.347DBRH-MHFL-TO8, Photodigm) is coupled through the optical fiber to monitor optical transmission during fabrication.

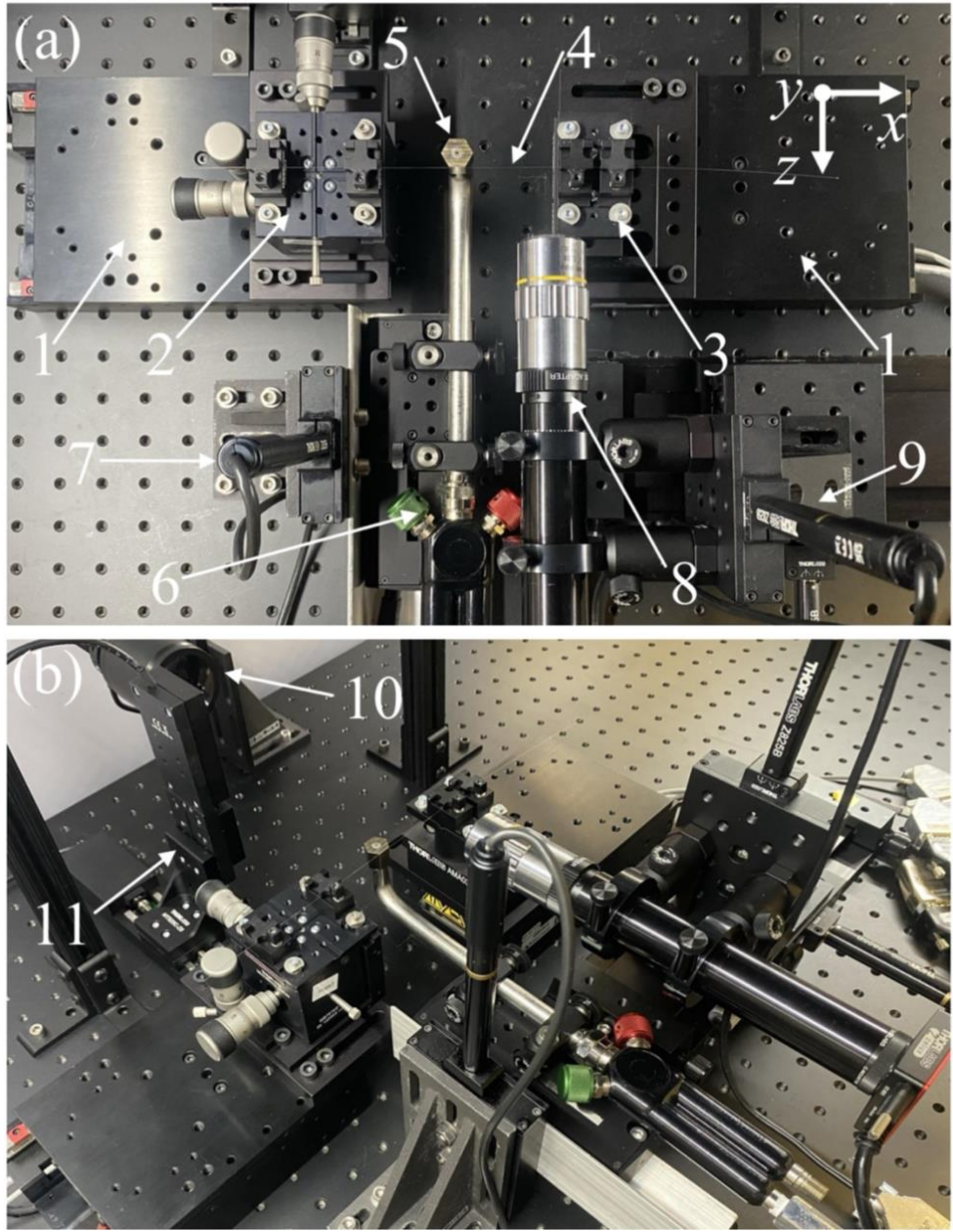

Fig. 2. Heat-and-pull Apparatus. (a) Top view and (b) 45° perspective view of the experimental setup. 1: Ultra-precision motorized 1-axis stages; 2: Optical fiber holders on a manual 3-axis stage; 3: Optical fiber holders on a stationary stage; 4: Single-mode optical fiber; 5: Torch tip; 6: Torch handle on a 2-axis motorized stage; 7: Horizontal digital microscope on a 2-axis motorized stage; 8: Vertical digital microscope on a 2-axis motorized stage; 9: Sample holder on a 2-axis motorized stage; 10: Light source.

We set the initial separation between the two pairs of optical fiber holders to ~3 cm to minimize alignment errors over the 100-mm travel range. As the stages scan, we use digital microscopes to measure the position of the single-mode optical fiber. Over the 100-mm travel range, the relative flatness and straightness errors are measured at $0 \pm 1$ $\mu$m and $2 \pm 1$ $\mu$m, respectively. Additionally, the relative yaw and pitch errors between the two stages are estimated by monitoring the position of a collimated laser beam reflected between the stages and are found to be negligible over the 100-mm travel range.

The single-mode optical fiber is heated and softened using an oxyhydrogen flame. Mixed hydrogen and oxygen gases are produced by electrolysis of distilled water using a low-cost generator. The mixed gas is first passed through two in-line particle filters (SS-4F-7, Swagelok) with a pore size of 7 $\mu$m and then regulated by a mass flow controller (GFCS-010007, Aalborg) with a flow range of 0-500 mL/min. The gas mixture is delivered through a torch handle (3H Hydrogen, National Torches) and combusted on a torch tip. The torch handle is mounted on a motorized 2-axis stage to align the flame with the single-mode optical fiber.

As shown in Fig. 3, three torch tips with different hole geometries are investigated: a commercial single-hole torch tip (OX-00, National Torches) with a diameter of 0.5 mm, and two custom-designed multi-hole torch tips with areas of 0.8×0.8 mm² and 1.9×1 mm², respectively. All the torch tips are made from stainless steel. Compared with single-hole torch tips, multi-hole torch tips provide a more homogeneous flame distribution, reduce the gas-flow pressure acting on the nanofiber, and mitigate the risk of combustion flashback [22, 24, 25, 27].

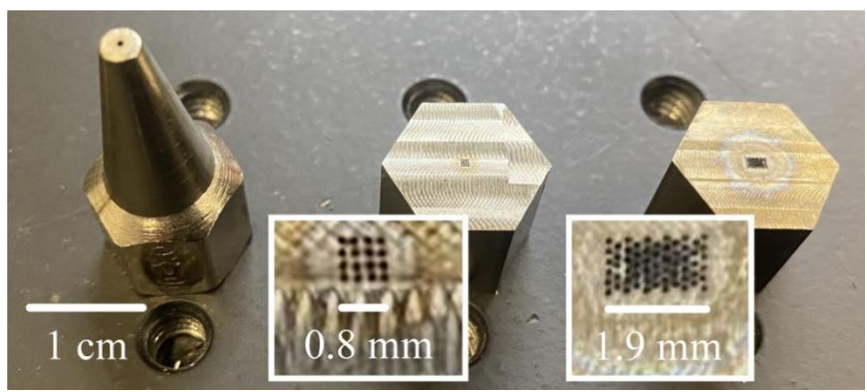

Fig. 3. Torch tips. Left: A single-hole torch tip with a diameter of 0.5 mm. Middle: A 16-hole torch tip. Right: A 61-hole torch tip. Each hole has a diameter of 0.15 mm. The inserted pictures are zoomed pictures of the holes.

The effective flame size of a torch tip is a critical parameter in the heat-and-pull process because it defines the local heating length and thus influences the resulting taper profile. As shown in Fig. 4, we experimentally determine the effective flame size for different torch tip geometries using a stationary-flame pulling method. The optical fiber is aligned with respect to the torch tip using the two digital microscopes. To ensure a stable flame while minimizing gas-flow pressure on the optical fiber, the oxyhydrogen flow rate is set to 110 mL/min, slightly above the minimum stable combustion threshold of 90 mL/min.

We investigate the influence of the fiber-flame separation distance on fabrication reliability. When the optical fiber is far from the flame, insufficient heating increases the probability of pulling fracture. Conversely, when the optical fiber is close, the bent fiber touches the torch tip during fabrication. Based on repeated trials, we determine an initial separation range of 0.2-2 mm, yielding stable and reproducible fabrication. Then, we further optimize the flame height for each torch tip by evaluating the repeatability of flame size measurements. For the 61-hole torch tip, consistent flame size is measured when the flame height is between 0.4 and 0.9 mm. Accordingly, we set the flame height to ~0.7 mm for subsequent experiments. With similar procedures, the flame heights of the single-hole and 16-hole torch tips are set to ~1.5 and ~0.7 mm, respectively.

During flame size measurements, the flame is held stationary while the optical fiber is symmetrically stretched in both axial directions at a constant relative pulling velocity of 0.1 mm/s. Prior to pulling, the initial optical fiber diameter $D_0$=125 μm is used to calibrate the horizontal digital microscope. As the optical fiber is elongated, the fiber diameter $D(x)$ is measured as a function of the axial extension $x$, as illustrated in Figs. 4 (b) and (c).

According to the exponential tapering law, the flame size $L_0$ is extracted by performing a linear fit to $\ln(D(x)/D_0)$ as a function of $x$, as shown in Fig. 4 (a). The flame sizes of the single-hole, 16-hole, and 61-hole torch tips are measured to be $0.36 \pm 0.01$ mm, $0.37 \pm 0.02$ mm, and $0.71 \pm 0.02$ mm, respectively.

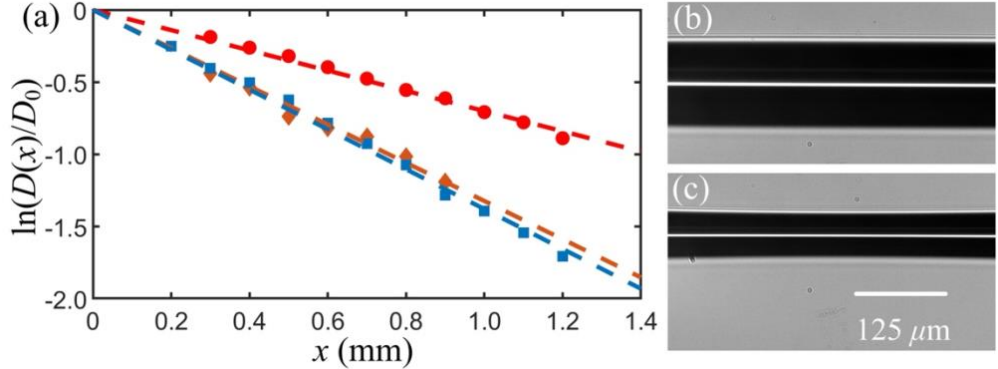

Fig. 4. Flame size measurements. (a) Normalized optical fiber diameter as a function of axial extension. Blue square: single-hole torch tip; Orange diamond: 16-hole torch tip; Red circle: 61-hole torch tip. Each data point represents the average of three measurements, and the error bars are smaller than the symbol size. (b) Optical microscope image of the optical fiber with an initial diameter of 125 μm before pulling. (c) Optical microscope image of the optical fiber with a reduced diameter of ~60 μm after an extension of 1.1 mm.

## 2.3 Software

We utilize custom software to design, fabricate, and characterize optical nanofibers. As illustrated in Fig. 5 (a), the software workflow consists of three sequential steps: taper and waist design, motion control generation, and fabrication and measurement.

To satisfy the adiabatic condition while minimizing the overall taper length, the taper profile is designed using a combination of two linear tapers and an exponential taper [37, 41]. Starting from the single-mode optical fiber, the first linear taper has an angle of 4 mrad and a length of ~5 mm. The second linear taper follows with a reduced angle of 3 mrad and a length of ~18 mm. An exponential taper then smoothly connects the second linear taper to the nanofiber waist. Fig. 5 (b) shows the profile with a nanofiber waist diameter of 200 nm and a waist length of 10 mm.

To translate the designed taper profile into executable stage trajectories, we employ an inverted iterative method [21, 22]. Given the target taper geometry, the stage motions are calculated by iteratively propagating backward from the nanofiber waist toward the single-mode fiber. At each iteration, the flame-scanning amplitude and axial elongation are determined based on the measured flame size and volume conservation, and are subsequently converted into synchronized position–velocity–time (PVT) trajectories for the two translation stages. In our experiment, the flame-scanning velocity is fixed at 2.5 mm/s, while the optical fiber pulling velocity is dynamically adjusted according to the local taper profile. Fig. 5 (c) illustrates the stage trajectories used to fabricate the designed taper profile shown in Fig. 5 (b).

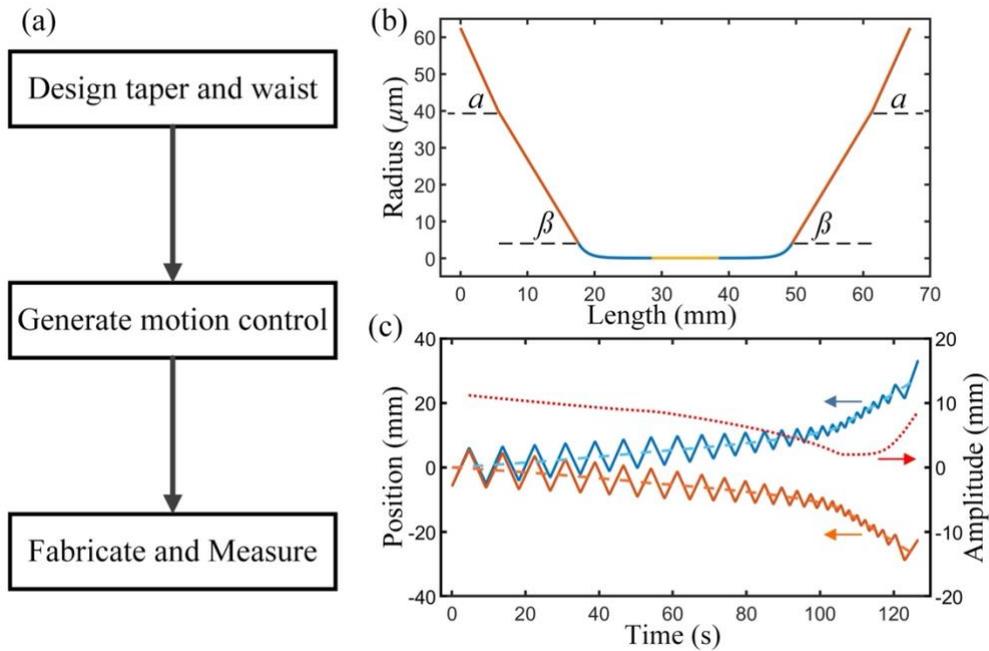

Fig. 5. Optical nanofiber fabrication software. (a) Flowchart of the heat-and-pull procedure. (b) A designed profile with a nanofiber waist of 200-nm diameter and 10-mm length (yellow), an exponential taper (blue), and two linear tapers with angles of $α=4$ mrad and $β=3$ mrad (orange). (c) Stage trajectories. The blue and orange solid curves represent the positions of the two translation stages, while the blue and orange dashed curves indicate the axial elongation of the optical fiber. The red dashed curve denotes the flame-scanning amplitude.

To implement the inverted iterative algorithm, we adopt the program developed by Hoffman [25]. Compared with implementations by Lützler [22] and Gouraud [26], Hoffman's program incorporates smooth acceleration and deceleration profiles for the translation stages, thereby suppressing mechanical transients and enhancing the reliability of nanofiber fabrication. We develop a LabVIEW program to execute the PVT files and integrate auxiliary functions to control the nanofiber fabrication, including adjusting the oxyhydrogen gas flow rate, positioning the torch tip relative to the optical fiber, observing the optical fiber diameter using digital microscopes, and monitoring optical transmission during fabrication.

We measure the fabricated nanofiber profiles using optical and electron microscopes. Fig. 6 shows two optical nanofibers with waist diameters of 200 nm and 400 nm and a waist length of 10 mm. To account for the additional fiber elongation during acceleration and deceleration of the translation stages, we correct the fiber profile shown in Fig. 5 (b) by applying a scaling coefficient and compare it with the measurements. In the taper regions, the diameters measured by optical microscopy closely follow the designed profile. High-resolution electron microscopy further verifies the uniformity and surface quality of the nanofiber waist. The agreement between the designed and measured profiles confirms the accuracy of the inverted iterative method and the reproducibility of the nanofiber fabrication process.

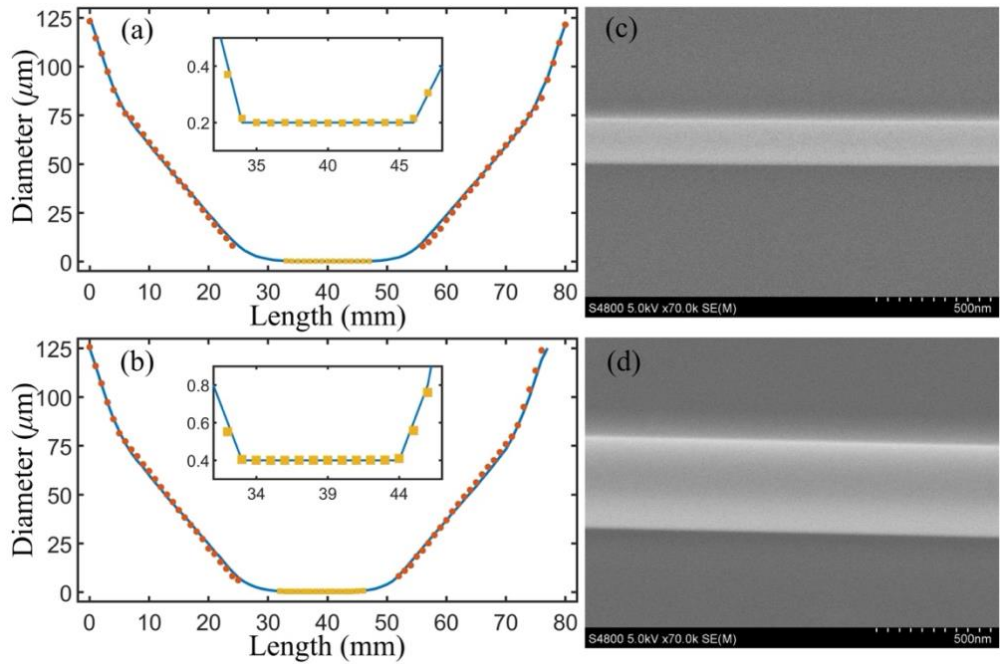

Fig. 6. Nanofiber profile measurements. (a) Optical nanofiber with a 200-nm waist diameter and a 10-mm waist length. (b) Optical nanofiber with a 400-nm waist diameter and a 10-mm waist length. The blue curves represent the designed nanofiber profile corrected by a scaling coefficient. Orange circles denote measurements from an optical microscope, and yellow squares denote measurements from an electron microscope. Each data point represents the average of three measurements, and the error bars are smaller than the symbol size. (c) Electron microscopy image of an optical nanofiber with a 200-nm waist diameter. (d) Electron microscopy image of an optical nanofiber with a 400-nm waist diameter.

## 3. Fabrication and Characterization of Optical Nanofibers

### 3.1 Optical Transmission Affected by Flame Shape

We investigate the influence of torch-tip flame geometry on fabrication stability and optical transmission. Three torch tips with different hole geometries are examined, as introduced in Section 2.2: a single-hole torch tip, a 16-hole torch tip, and a 61-hole torch tip. These torches produce flames with different effective sizes and gas-flow pressure, which can affect both heating uniformity and mechanical stability during pulling. To evaluate their performance, optical nanofibers with a 300-nm waist diameter are fabricated with two representative waist lengths of 1 and 50 mm. Optical transmission is monitored throughout the fabrication process, and short-time Fourier transform (STFT) analysis is applied to the transmission traces to reveal their frequency components.

Figure 7 shows the experimental results. For the single-hole torch tip, Figs. 7 (a) and (b) show the transmission trace and corresponding STFT spectrum for the fabrication of a nanofiber with a 1-mm waist length, while Figs. 7 (c) and (d) present the results for a 50-mm waist length. In both cases, significant transmission oscillations appear during the later stage of fabrication as the fiber diameter decreases. These oscillations arise from the beating of multiple guided optical modes as the optical fiber vibrates in the gas flow. The STFT spectra show that the beating frequency increases until the fiber diameter reaches the cutoff condition for the higher-order optical modes.

For the 16-hole torch tip, the results are shown in Figs. 7 (e) and (f) for the 1-mm waist length and in Figs. 7 (g) and (h) for the 50-mm waist length. Compared with the single-hole torch, transmission becomes more stable, although occasional transient fluctuations persist. The STFT spectra show reduced amplitude of the beating frequency, suggesting that the distributed flame produced by the 16-hole torch mitigates gas-flow-induced perturbations.

The most stable fabrication is achieved with the 61-hole torch tip. Figs. 7 (i) and (j) show the transmission and STFT results for the 1-mm waist length, while Figs. 7 (k) and (l) for the 50-mm waist length. In both cases, the transmission remains above 99.9% throughout the fabrication process with only minimal fluctuations. The STFT spectra exhibit negligible multi-mode beating signatures.

For optical nanofibers with identical waist dimensions, fabrication with the single-hole and 16-hole torch tips takes longer than with the 61-hole torch tip. This difference arises because the effective flame size of the 61-hole torch tip is about twice that of the single-hole and 16-hole torch tips. The larger flame size enables faster fabrication and reduces the risk of dust accumulation during pulling, though it also limits the minimum achievable waist length.

These results demonstrate that the flame shape influences fabrication stability. Generally, torch tips that produce a relatively large and uniform heating area at low gas-flow pressure are preferred for optical nanofiber fabrication. Among the tested configurations, the 61-hole torch tip provides the most homogeneous flame and the highest transmission stability. Therefore, the 61-hole torch tip is adopted for all subsequent experiments presented in this work.

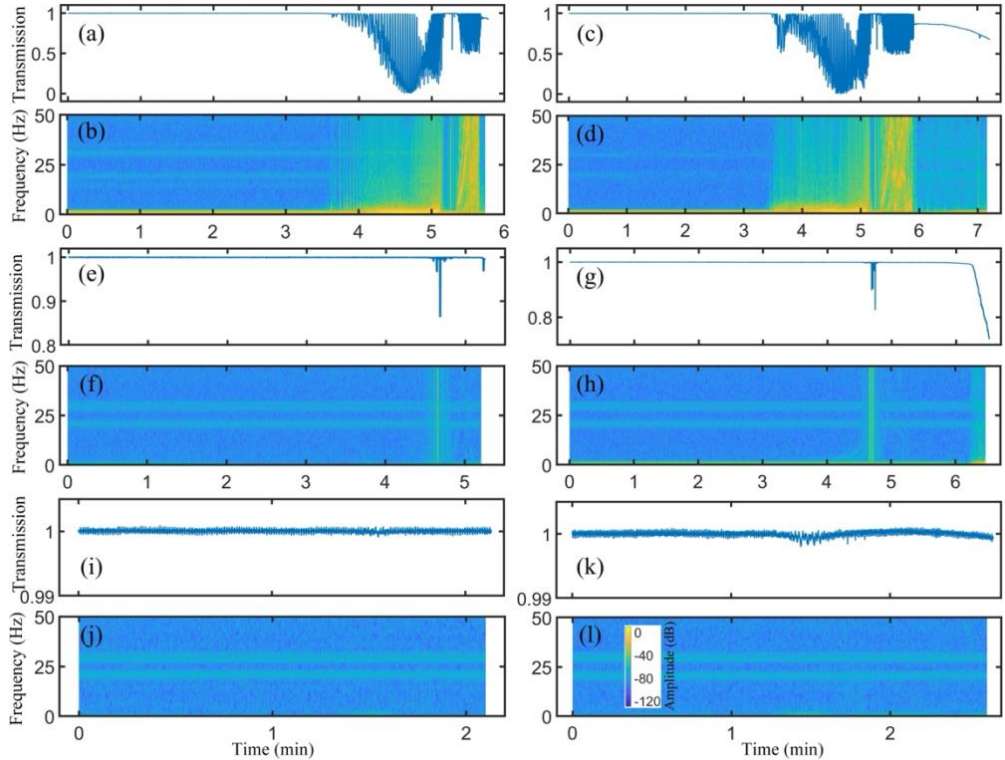

Fig. 7. Optical transmission during nanofiber fabrication and the corresponding short-time Fourier transform (STFT) analysis for different torch tips. All nanofibers have a 300-nm waist diameter. (a) and (b) Transmission and STFT during fabrication using a single-hole torch tip with a 1-mm waist length; (c) and (d) corresponding results for a 50-mm waist length. (e) and (f) Results for a 16-hole torch tip with a 1-mm waist length; (g) and (h) corresponding results for a 50-mm waist length. (i) and (j) Results for a 61-hole torch tip with a 1-mm waist length; (k) and (l) corresponding results for a 50-mm waist length.

## 3.2 Optical Transmission Affected by Optical Nanofiber Waist and Length

Optical transmission is an important metric for evaluating the quality of nanofiber fabrication. In addition, real-time monitoring of transmission during fabrication provides information on the mechanical stability of the optical fiber, revealing potential fabrication fractures or deviations from the designed profile. To examine the influence of waist diameters and lengths on optical transmission, we fabricate optical nanofibers with waist diameters of 180, 200, 250, and 300 nm and waist lengths of 1 and 50 mm.

Figure 8 (a) shows the optical transmission during the fabrication of optical nanofibers with a 1-mm waist length. Fabrication begins at 0.5 min and ends at around 2.6 min. Note that thinner-waist nanofibers cost slightly longer fabrication times than thicker ones, due to the additional pulling required to achieve smaller diameters. For waist diameters of 250 and 300 nm, the optical transmission remains above 99.9% throughout the entire fabrication process.

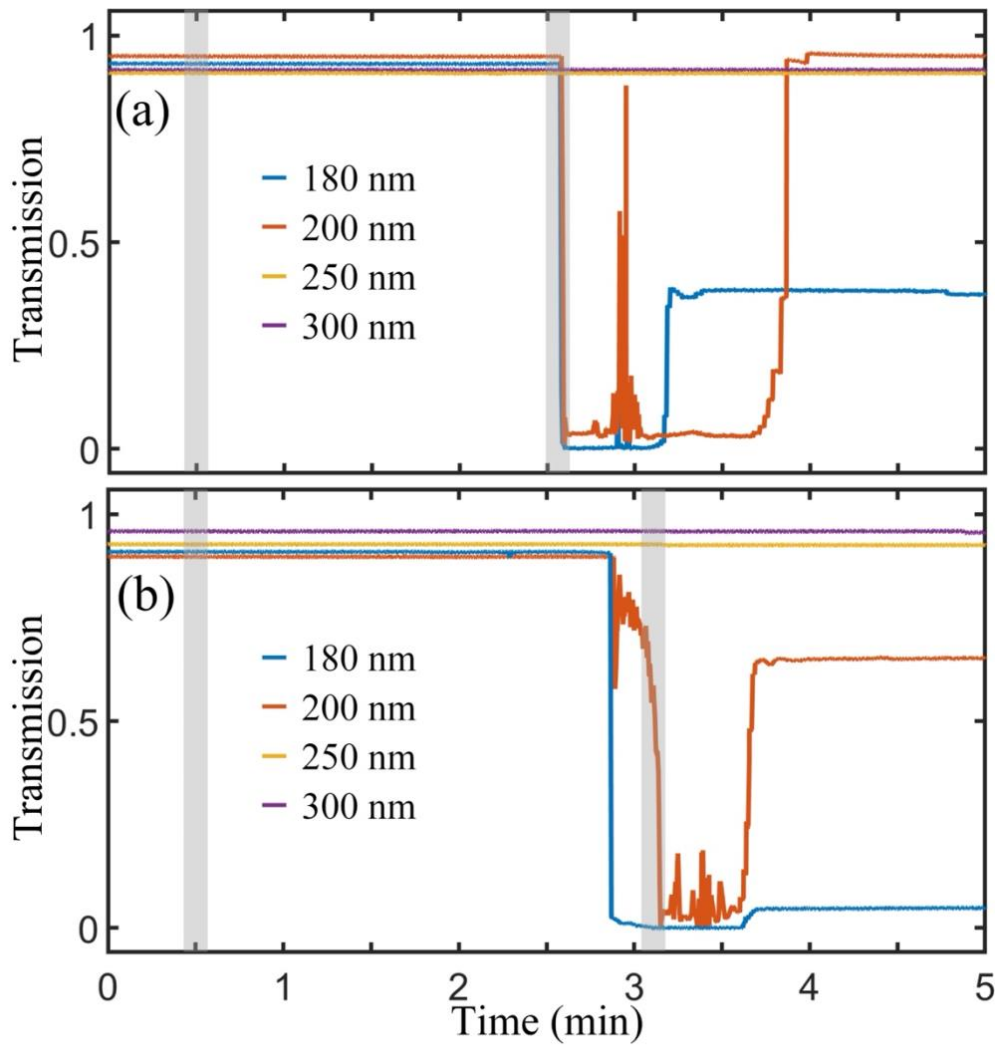

Fig. 8. Optical transmission during fabrication for different nanofiber waist diameters and lengths. (a) Transmission traces for a waist length of 1 mm. (b) Transmission traces for a waist length of 50 mm. The blue, orange, yellow, and purple curves correspond to waist diameters of 180, 200, 250, and 300 nm, respectively. The shaded regions indicate the start and end of the fabrication process.

For smaller waist diameters of 180 and 200 nm, the transmission drops abruptly at the end of the fabrication process. When pulling stops, the transmission of the 180-nm nanofiber falls to 2%, while that of the 200-nm nanofiber decreases to 5%. We attribute a portion of the transmission loss to residual bending of the nanofiber induced during fabrication. To recover the transmission, we apply a small post-fabrication extension by translating one stage in steps of 0.02 mm. As the optical nanofiber is gradually extended, the transmission increases. For the 180-nm nanofiber, we begin the extension at 3.1 min, and the transmission reaches 40% after an extension of 0.12 mm. For the 200-nm nanofiber, the extension starts at 3.6 min, and the transmission is restored to above 99.9% after an extension of 0.16 mm. Further extension produces little change in transmission. The transmission fluctuations observed between the end of fabrication and the start of post-fabrication extension are attributed to vibrations of the bent nanofibers caused by ambient airflow.

Figure 8 (b) presents the optical transmission of optical nanofibers with a 50-mm waist length. Fabrication starts at 0.5 min and ends at around 3.1 min. Similar to the 1-mm case, nanofibers with waist diameters of 250 and 300 nm maintain transmission above 99.9% throughout fabrication. For the 180-nm nanofiber, transmission drops to zero about 13 s before fabrication ends. A post-fabrication extension of 0.16 mm recovers the transmission to 10%. For the 200-nm nanofiber, transmission begins to decrease gradually about 15 s before the end of fabrication and then drops sharply to 3% at the stopping point. A post-fabrication extension of 0.16 mm increases the transmission to 75%.

To evaluate the mechanical limits of post-fabrication extension, we gradually extend the nanofibers while monitoring optical transmission until fracture occurs. Optical nanofibers with 1-mm waist length fail at extensions of around 0.4 mm, whereas those with 50-mm waist length sustain extensions of up to about 2.8 mm before breaking. For comparison, we also test nanofibers with a 10-mm waist length, which exhibit an intermediate failure extension of around 0.8 mm. In contrast, when bending the nanofibers by moving the stages closer together, nanofibers of all tested waist diameters and lengths tolerate bending radii on the order of tens of millimeters without breaking.

These observations suggest that waist diameters and lengths influence the optical transmission and mechanical robustness. Under comparable fabrication conditions, optical nanofibers with a 1-mm waist length show more resilient optical transmission than those with a 50-mm waist length, and thicker waist diameters tend to exhibit better transmission than thinner ones. Selecting sufficient waist diameters and lengths is important for balancing fabrication yield and optical transmission across different applications.

*3.3 Optical Transmission Affected by Dust*

Surface contamination is a major factor that degrades optical transmission in optical nanofibers. Owing to their subwavelength diameters, optical nanofibers are sensitive to surface inhomogeneities, which introduce scattering loss and compromise mechanical stability during fabrication. Dust can be present prior to fabrication if cleaning is insufficient, and it can also accumulate during and after fabrication under ambient conditions.

To minimize contamination prior to fabrication, a standardized cleaning protocol is implemented. First, a 3–5 cm section of the optical fiber coating is removed using a stripping tool (T06S13, Thorlabs). Second, the exposed fiber is wiped with a lint-free wipe (TCW604, Thorlabs) to remove residual debris, and then cleaved with a manual cleaver (S326A, Fitel) for subsequent fusion splicing. Third, the prepared fiber, with a bare length of 2–4 cm, is spliced to another fiber of the same type with a bare length of approximately 0.5–1 cm using a fusion splicer (S179A, Fitel). After splicing, the total bare fiber length is approximately 2–5 cm, with the splice point located near one end of the coated optical fiber. Fourth, the bare fiber section is gently cleaned with a lint-free wipe containing precision cleaning fluid (FCS3, Thorlabs), followed by a final dry lint-free wipe. Finally, the optical fiber is mounted on the fabrication rig and inspected under a microscope by focusing on the front surface and the upper and lower edges. If dust is observed on the optical fiber, the cleaning step is repeated. The splice point is positioned outside the flame-scanning region and therefore does not affect fabrication quality.

Figure 9 (a) compares transmission traces for optical nanofibers with a 300-nm waist diameter and a 30-mm waist length fabricated under different cleaning conditions. The shaded region indicates the start and end of the fabrication process. The well-cleaned optical fiber, shown in Fig. 9 (b), maintains nearly constant transmission exceeding 99.9% throughout fabrication. In contrast, optical fibers with insufficient cleaning, shown in Fig. 9 (c), exhibit significant transmission fluctuations once fabrication begins. After fabrication, the transmission is lower than that of the well-cleaned optical fiber. In severe cases, the transmission abruptly drops to zero near the end of the fabrication process, indicating optical fiber fracture. These observations suggest that dust or oil on the optical fiber surface interacts with the flame during fabrication, degrading the nanofiber surface quality and increasing both optical loss and mechanical fragility.

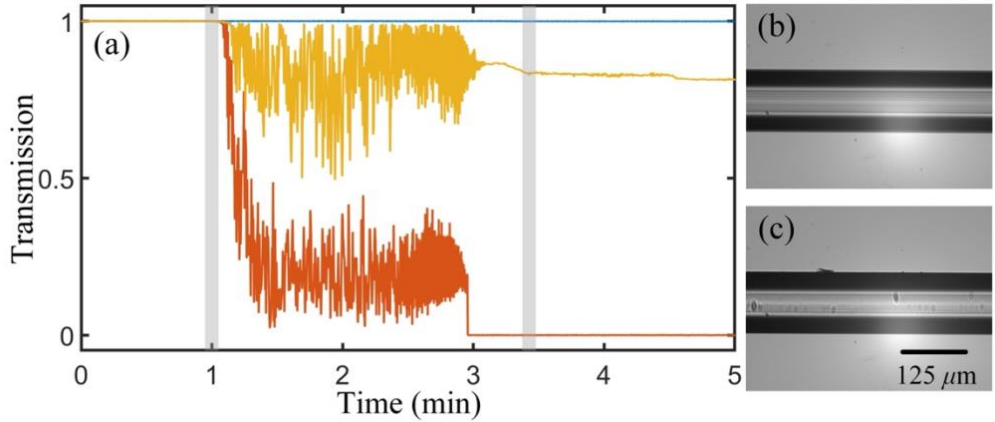

Fig. 9. Optical transmission during fabrication for nanofibers with a 300-nm waist diameter and a 30-mm waist length under different cleaning conditions. (a) Transmission traces for a well-cleaned fiber (blue) and insufficiently cleaned fibers (yellow and orange). The shaded region indicates the start and completion of fabrication. (b) Microscope image of a well-cleaned optical fiber. (c) Microscope image of a contaminated optical fiber.

Even when fabrication begins with a properly cleaned optical fiber, transmission will degrade over time due to dust accumulation in ambient air. Fig. 10 (a) presents long-term transmission measurements for optical nanofibers with a 300-nm waist diameter and waist lengths of 1, 10, 30, and 50 mm. Fabrications start at 1 min and complete at around 3.3 min, and the transmission is monitored for 2 hours. These measurements are performed in a standard laboratory environment rather than in a cleanroom. The fabrication setup is enclosed within an aluminum housing to reduce airflow and dust deposition. Under these conditions, nanofibers with waist lengths of 1, 10, and 30 mm maintain transmission above 85% for more than 1 hour, while those with a 50-mm waist length maintain transmission above 50% for 0.5 hour. This duration provides sufficient time to conduct experiments or transfer the nanofiber to a more protected environment.

The 1-mm nanofiber maintains transmission exceeding 99.9%. In contrast, nanofibers with longer waist exhibit transmission decay. Step-like drops are occasionally observed, consistent with discrete dust attachment events. Gradual decreases suggest continuous accumulation of smaller particles. Fluorescence images in Fig. 10 (b) and (c) confirm that attached dust particles produce localized scattering and thus induce transmission loss. The stronger degradation observed with longer waist lengths is attributed to their greater exposed surface area, which increases the probability of dust deposition.

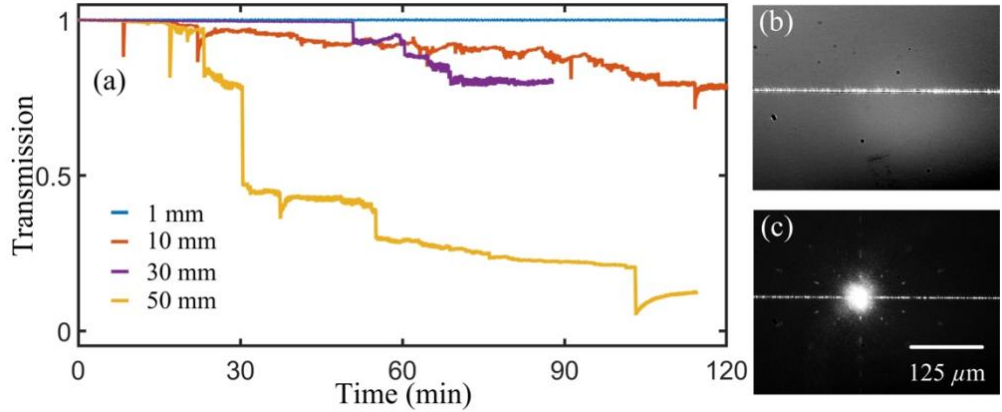

Fig. 10. Long-term optical transmission drift for nanofibers with a 300-nm waist diameter and different waist lengths. (a) Transmission traces for waist lengths of 1 mm (blue), 10 mm (orange), 30 mm (purple), and 50 mm (yellow). Fabrication starts at 1 minute and completes at around 3.3 minutes. (b) Fluorescent image of a clean nanofiber. (c) Fluorescent image of a nanofiber with an attached dust particle producing localized scattering.

These results show that dust contamination is a dominant mechanism affecting both fabrication stability and long-term transmission performance. Careful cleaning prior to fabrication and minimizing environmental exposure afterward are essential for achieving high, stable optical transmission, particularly for nanofibers with a thin waist diameter and a long waist length.

## 4. Conclusion

We present an optical nanofiber fabrication system based on the heat-and-pull fabrication process, including the hardware implementation, control software, and system performance. By investigating the effects of flame geometry, nanofiber waist dimensions, and environmental contamination, we identify several factors that affect fabrication stability and optical transmission performance.

The geometry of the heating flame influences the fabrication time and stability of optical nanofibers. Compared with the single-hole and 16-hole torch tips, the 61-hole torch tip has the largest effective flame size of ~0.7 mm, producing a more homogeneous flame and reducing gas-flow-induced vibration of the nanofiber. Consequently, fabrication becomes faster and yields higher optical transmission over a wide range of nanofiber waist diameters and lengths. We further compare the optical transmission of nanofibers with different waist diameters and lengths, demonstrating transmission above 99.9% for a 50-mm waist length with a waist diameter as small as 250 nm. For a short waist length of 1 mm, optical transmission above 99.9% is achieved with waist diameters as small as 200 nm. In addition, we develop a preparation protocol that includes optical fiber splicing and cleaning to minimize fabrication failures and transmission degradation caused by surface contamination. Although optical transmission gradually decays over time due to dust accumulation, nanofibers with a waist diameter of 300 nm and waist lengths of 1-30 mm maintain transmission above 85% for more than 1 hour in a typical laboratory environment when fabricated in an enclosed area. These results provide practical guidance for constructing custom optical nanofiber fabrication systems and reliably producing low-loss optical nanofibers for optics and atomic physics experiments.

**Funding.** National Science Foundation (2316595, 2328663).

**Acknowledgment.** We thank Kristina Smsky, Chloe Kim, Zara Zhao, and Alex Radak for their contributions to the experiment. G. S. thanks the support of the Dean's Dissertation Fellowship from Rutgers Graduate School-Newark.

**Disclosures.** The authors declare that they have no conflicts of interest.

**Data availability.** Data underlying the results presented in this paper are not publicly available at this time but may be obtained from the authors upon reasonable request.